\documentclass[prl,aps,a4paper,showpacs,,superscriptaddress,twocolumn]{revtex4}
\usepackage{amsfonts}
\usepackage{amsmath}
\usepackage{graphicx}
\usepackage{dcolumn}
\usepackage{bm}

\begin{document}
\title{The generalized Gibbs ensemble as a pseudo-initial state: its predictive power revealed in a second quench}
\author{J.~M. Zhang}
\affiliation {Institute of Physics, Chinese Academy of
Sciences, Beijing 100080, China}
\author{F.~C. Cui}
\affiliation {Institute of Physics, Chinese Academy of
Sciences, Beijing 100080, China}
\author{Jiangping Hu}
\affiliation {Institute of Physics, Chinese Academy of
Sciences, Beijing 100080, China} \affiliation {Department
of Physics, Purdue University, West Lafayette, IN 47906}

\begin{abstract}
The generalized Gibbs ensemble has been shown to be
relevant in the relaxation of a completely integrable
system subject to a quantum quench, in the sense that it
accurately predicts the steady values of some physical
variables. We proceed to further question its relevance by
giving the quenched system a second quench. The concern is
whether the generalized Gibbs ensemble can also accurately
predict the relaxed system's response to the second quench.
Two case studies with the transverse Ising model and the
hard-core bosons in one dimension yield an affirmative
answer. The relevance of the generalized Gibbs ensemble in
the non-equilibrium dynamics of integrable systems is then
greatly strengthened.
\end{abstract}

\maketitle

Recently, non-equilibrium dynamics of many-body systems has
attracted a lot of attention \cite{Polkovnikov}. One common
concern is whether an initially out-of-equilibrium system
can thermalize to behave like a textbook Gibbs ensemble,
and how integrability \cite{notion,weiss} or
non-integrability of the system will affect its relaxation
dynamics. An important achievement on this issue is
identification of the relevance of the generalized Gibbs
ensemble (GGE) in the relaxation dynamics of a completely
integrable system \cite{rigol07}. The so called generalized
Gibbs ensemble is constructed according to the principle of
maximum entropy \cite{jaynes} while taking into account all
the constants of motion, whose values are determined by the
initial state. With the same philosophy behind the
construction, it is a natural counterpart of the usual
Gibbs ensembles for a non-integrable system. So far, the
GGE has been found to predict correctly the asymptotic
values of physical variables in a variety of integrable
systems
\cite{Polkovnikov,cardy07,rigol07,cazalilla,kollar08,Calabrese11,mussardo,caneva,cazalilla11,limitation}.
%In contrast, it is interesting that, for many
%non-integrable models \cite{kollath,zjm,manmana,hastings},
%the usual canonical or grand-canonical ensemble often can
%not do the same job.

The fact that asymptotically, the true, constantly evolving
system agrees well with the GGE on the physical quantities
is definitely a non-trivial and pleasant one. However, one
should not be content with this fact only. Our daily
experience in the (mostly non-integrable) macroscopic world
is that, if a system relaxes to some steady state, it
relaxes in the sense that not only its static properties
(i.e. values of the physical quantities) but also its
dynamical properties agree with the steady state. To be
specific, the system should respond to later perturbations
as if it were indeed in the steady state. Therefore, it is
necessary to check whether the GGE has this merit. If so,
it surely adds to the relevance of the GGE in the
non-equilibrium dynamics of a completely integrable system.
It would mean that the true system is hardly
distinguishable from the GGE neither by static nor
dynamical criterions, and it would be fair to say the
system has thermalized as much as possible.

Motivated by this problem, we have studied the transverse
Ising model and the hard-core bosons in one-dimension
(which can also be mapped to the XX model) individually.
The two models are integrable and both have been shown to
admit a GGE account of their asymptotic behaviors after a
quantum quench. Here our idea is to give them a second
quench when they have reached the steady phase \cite{zjm}.
The concern is whether they will respond as if the systems
were in the GGE states. The result turns out to be the
case.

\textit{Transverse Ising model}.---The Hamiltonian of the
model is $H(g)=-\sum_{l=1}^N \left( \sigma_l^x
\sigma_{l+1}^x - g \sigma_l^z \right)$, where
$\sigma_l^{x,z}$ are Pauli matrices acting on a $1/2$-spin
at site $l$. Here periodic boundary condition is assumed
and $N$ is an even integer large enough. Below, quenches of
the system correspond to changing the value of $g$
(strength of the transverse magnetic field) suddenly. We
will consider a double quench scenario. Initially the value
of $g$ is $g_0$ and the system is in its ground state
$|G_0\rangle$. Then the value of $g$ is changed
successively to $g_1$ and $g_2$.

Under the Jordan-Wigner transform $(\sigma_l^x+ i
\sigma_l^y, \sigma_l^x- i \sigma_l^y)/2=(a_l^\dagger, a_l)
\exp(-i \pi \sum_{r=1}^{l-1} a_r^\dagger a_r)$,
$\sigma_l^z= 2 a_l^\dagger a_l - 1$, where $a_l^\dagger$
and $a_l$ are fermionic operators, the Hamiltonian is
rewritten as $ H(g)= -\sum_{l=1}^N [ \left( a_l^+ a_{l+1}^+
+ a_l^+ a_{l+1} +h.c.\right) - 2g  a_l^\dagger a_l ] $,
with a constant term dropped \cite{lieb}. Note that here
the boundary condition is anti-periodic \cite{supple}.
Taking the Fourier transform $b_k=\frac{1}{\sqrt{N}} \sum_l
e^{i 2\pi k l/N } a_l$, with
$k=-N/2+1/2,\cdots,-1/2,1/2,\cdots,N/2-1/2$ so as to comply
with the anti-periodic boundary condition, we can rewrite
the Hamiltonian as ($\phi_k= 2\pi k/N$)
\begin{equation}\label{h}
 H(g)= \sum_k \left[2(g-\cos\phi_k)b_k^\dagger b_k -i \sin \phi_k (b_{-k}b_k + b_{-k}^\dagger b_k^\dagger)
 \right]. \nonumber
\end{equation}
It is ready to verify that $b_k$ and $b_{-k}^\dagger$ are
coupled in their equations of motion and this suggests the
Bogoliubov transformation $\eta_k= u_k b_k + i v_k
b_{-k}^\dagger$. With $\varepsilon_k =2 \sqrt{1+g^2 - 2g
\cos \phi_k} \geq 0$,
$(\cos\theta_k,\sin\theta_k)=2(g-\cos\phi_k,\sin\phi_k)/\varepsilon_k$,
and $(u_k,v_k)=(\cos \frac{\theta_k}{2}, \sin
\frac{\theta_k}{2})$, the Hamiltonian is finally
diagonalized as $H= \sum_k \varepsilon_k \eta_k^\dagger
\eta_k $. Here again the constant term is dropped. Note
that $u_k$, $v_k$, $\theta_k$, and $\varepsilon_k$ all
depend on $g$. The dependence will be displayed explicitly
when necessary.

We are interested in the correlation functions $\langle
\sigma_i^x \sigma_j^x \rangle$, $\langle \sigma_i^z
\sigma_j^z \rangle$, and the transverse magnetization
$\langle M_z \rangle\equiv \langle \sum_l \sigma_l^z
\rangle =\langle \sum_k (2 b_k^\dagger b_k -1) \rangle$.
Here the expectation values may be taken with respect to
various states as shown below. Introducing $A_l=
a_l^\dagger + a_l$ and $B_l= a_l^\dagger- a_l$, we can
rewrite them as $\langle \sigma_i^x \sigma_j^x \rangle
=\langle B_iA_{i+1} B_{i+1} \cdots A_{j-1} B_{j-1} A_j
\rangle$ and $\langle \sigma_i^z \sigma_j^z \rangle=
\langle B_iA_{i} B_{j}A_j \rangle $ \cite{lieb}. These
forms allow us to use Wick's theorem to do the calculation.
The correlation functions will be decomposed into sums of
products of the basic correlators $\langle A_lA_m \rangle$,
$\langle B_lB_m \rangle$, and $\langle B_lA_m \rangle$.

The initial state $|G_0\rangle $ is defined as
$\eta_k(g_0)|G_0\rangle=0$ for all $k$, or explicitly,
$|G_0\rangle\propto \prod_k \eta_k(g_0)|\psi\rangle $ where
$|\psi\rangle$ can be an arbitrary state as long as
$\eta_k(g_0)|\psi\rangle\neq 0$. After the first quench of
changing $g$ from $g_0$ to $g_1$ at $t=0$, we have $\langle
G_0 | A_l(t) A_m(t) | G_0 \rangle\rightarrow \delta_{lm}$
for $t$ large enough \cite{recurrence,sengupta}, and
similarly $\langle G_0 | B_l(t) B_m(t) | G_0
\rangle\rightarrow -\delta_{lm}$ for $t$ large enough. But
$\langle G_0 | B_l(t) A_m(t)  | G_0 \rangle \rightarrow
G_{l,m}^{(1)}$, which has the value of
\begin{equation}\label{albm}
    G_{l,m}^{(1)}=
-\frac{1}{N} \sum_k e^{i\phi_k(m-l)+i
\theta_k(g_1)}\cos(\Delta\theta_k^{10}).
\end{equation}
Here and hereafter $\Delta\theta_k^{ij}\equiv
\theta_k(g_i)-\theta_k(g_j)$. Thus for $t$ large enough,
$\langle G_0| \sigma_i^x(t) \sigma_j^x(t) |G_0\rangle
\rightarrow C_{ij}^x$:
\begin{equation}\label{cijx}
C_{ij}^x= \text{Det}\left( \begin{array}{cccc}
G_{i,i+1}^{(1)} & G_{i, i+2}^{(1)} & \cdots & G_{i,j}^{(1)} \\
G_{i+1,i+1}^{(1)} & G_{i+1, i+2}^{(1)} & \cdots & G_{i+1,j}^{(1)} \\
\cdots & \cdots & \cdots & \cdots \\
G_{j-1,i+1}^{(1)} & G_{j-1, i+2}^{(1)} & \cdots &
G_{j-1,j}^{(1)}
\end{array}  \right),
\end{equation}
and
\begin{equation}\label{sigmazz}
\langle G_0| \sigma_i^z(t) \sigma_j^z(t) |G_0\rangle
\rightarrow C_{ij}^z = G_{i,i}^{(1)} G_{j,j}^{(1)} -
G_{i,j}^{(1)} G_{j,i}^{(1)}.
\end{equation}
As for the transverse magnetization, $ \langle \Psi_0
|M_z(t)|\Psi_0 \rangle $ has the asymptotic value of
\begin{equation}\label{transemag1}
  M_z^{(1)} = -\sum_k \cos\theta_k(g_1)
   \cos(\Delta\theta_k^{10}).
\end{equation}
On the other hand, from $g_0$ to $g_1$, the (first) GGE
density matrix is defined as
\begin{equation}\label{firstgge}
\rho_{gge1}= \frac{1}{Z_1}\prod_k \exp
\left(-\lambda_k^{(1)} \eta_k^\dagger(g_1) \eta_k(g_1)
\right),
\end{equation}
with the Lagrange multiplier $\lambda_k^{(1)}$ determined
by the condition $\langle G_0 | \eta_k^\dagger(g_1)
    \eta_k(g_1) |G_0 \rangle = tr(\eta_k^\dagger(g_1)
    \eta_k(g_1) \rho_{gge1})$, and $Z_1=\prod_k (1+e^{-\lambda_k^{(1)}})$. It can be
verified that $\langle A_lA_m \rangle_{gge1}= -\langle B_l
B_m \rangle_{gge1}= \delta_{lm}$, and $\langle B_l A_m
\rangle_{gge1} = G_{l,m}^{(1)}$. Here the subscript means
averaging over $\rho_{gge1}$. Thus the basic correlators
are of the same values with respect to the GGE density
matrix $\rho_{gge1}$ and the evolving state
$e^{-iH(g_1)t}|G_0\rangle$ for $t$ large enough. This fact
then indicates that the asymptotic values of the
correlation functions (\ref{cijx}) and (\ref{sigmazz}) can
be recovered with the GGE. Likewise, the asymptotic value
of the transverse magnetization (\ref{transemag1}) is
exactly predicted by the GGE, i.e., $M_z^{(1)} =tr(M_z
\rho_{gge1})$ \cite{Polkovnikov,supple}.

Now consider giving the system a second quench, i.e.,
changing the value of $g$ from $g_1$ to $g_2$ at some time
$t=t_1$. It is tedious but straightforward to show that at
the time of $t=t_1+t_2$, for large $t_2$ \cite{supple},
$\langle G_0 | A_l(t) A_m(t) | G_0 \rangle \simeq
\delta_{lm} +$oscillating terms depending on $t_1$, and
similarly $\langle G_0 | B_l(t) B_m(t) |G_0 \rangle \simeq
-\delta_{lm}+ $oscillating terms depending on $t_1$.
However, $\langle G_0 |  B_l(t)A_m(t) |G_0 \rangle \simeq
G_{l,m}^{(2)}+$oscillating terms depending on $t_1$, where
\begin{eqnarray}\label{bmal3}
   G_{l,m}^{(2)} = -\frac{1}{N} \sum_k e^{i\phi_k(m-l)+i
\theta_k(g_2)}\cos(\Delta\theta_k^{10})\cos(\Delta\theta_k^{21}).
\end{eqnarray}
As for the transverse magnetization, $\langle
G_0|M_z(t)|G_0\rangle\rightarrow M_z^{(2)} +$oscillating
terms depending on $t_1$, with
%\begin{eqnarray} \label{transemag2}
%  M_z^{(2)} &=& -\sum_k \cos\theta_k(g_2) \cos(\theta_k(g_1) -\theta_k(g_0)) \nonumber \\
%&& \quad\quad\quad \times\cos(\theta_k(g_2)
%-\theta_k(g_1)).
%\end{eqnarray}
\begin{eqnarray} \label{transemag2}
  M_z^{(2)} =-\sum_k \cos\theta_k(g_2) \cos(\Delta\theta_k^{10})\cos(\Delta\theta_k^{21}).
\end{eqnarray}
The oscillating terms depending on $t_1$ consist of $O(N)$
components of different non-zero frequencies and thus they
virtually vanish for $t_1$ large enough. Therefore, for
$t_1$ and $t_2$ large enough, the correlation functions
$\langle G_0|\sigma_i^x(t) \sigma_j^x(t)|G_0 \rangle$ and
$\langle G_0| \sigma_i^z(t) \sigma_j^z (t) |G_0\rangle$
have the same form as Eqs.~(\ref{cijx}) and (\ref{sigmazz})
but with $G_{m,l}^{(1)}$ replaced by $ G_{m,l}^{(2)}$, and
$\langle G_0|M_z(t)|G_0\rangle$ has the value of
$M_z^{(2)}$.

\begin{figure}[tb]
\includegraphics[ width= 0.36\textwidth,bb= 18   188   587   615]{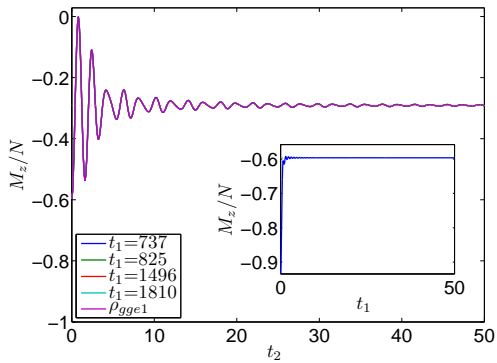}
\caption{(Color online) Evolution of the transverse
magnetization $M_z$ after the second quench. The parameters
are $(N,g_0,g_1,g_2)= (10000,2,1,0.2)$. All the lines, with
the ``initial'' state being the (first) generalized Gibbs
ensemble (GGE) density matrix $\rho_{gge1}$ or $e^{-i
H(g_1) t_1} |G_0\rangle$, collapse into one. Here the
values of $t_1$ are chosen randomly from $[500,2500]$. The
horizontal dotted line indicates the predicted asymptotic
value (\ref{transemag2}). The insert shows the time
evolution of $M_z$ after the first quench.\label{fig1}}
\end{figure}

On the other hand, if the second quench is imposed on the
first GGE density matrix $\rho_{gge1}$, we have the same
asymptotic behaviors of the basic correlators and the
transverse magnetization for large $t_2$. That is, $\langle
A_l(t_2) A_m(t_2) \rangle=-\langle B_l(t_2) B_m(t_2)
\rangle\simeq \delta_{lm}$, $\langle B_m(t_2) A_l(t_2)
\rangle\simeq G_{m,l}^{(2)} $, and $\langle M_z(t_2)
\rangle \simeq M_z^{(2)}$ \cite{supple}. Here
$(A_l(t_2),B_l(t_2),M_z(t_2))=e^{iH(g_2)t_2}(A_l,B_l,M_z)
e^{-iH(g_2)t_2}$ and the average is taken over
$\rho_{gge1}$. We see that the transverse magnetization as
well as the basic correlators possess the same asymptotic
values regardless of the initial state being $
e^{-iH(g_1)t_1}|G_0 \rangle$ or $\rho_{gge1}$. The latter
fact implies that the correlation functions have the same
property. However, it is not only the asymptotic values
that can be accurately reproduced by using $\rho_{gge1}$ as
a substitute for $e^{-iH(g_1)t_1}|G_0 \rangle$. In
Fig.~\ref{fig1}, the transient dynamics of $M_z$ after the
second quench is shown. There we see that as long as $t_1$
is large enough, the relaxation dynamics of $M_z$ (the
correlation functions have the same property; see the
supplementary material) is independent of $t_1$ and can be
reproduced by $\rho_{gge1}$ even to minute details.
Therefore, as long as $t_1$ is large enough, or as long as
the second quench comes when the system has equilibrated to
agree with the first GGE $\rho_{gge1}$ after the first
quench, the model reacts as if it were indeed in the GGE
state $\rho_{gge1}$. That is, the GGE density matrix
$\rho_{gge1}$ can serve as a pseudo-initial state to the
second quench.

Finally, for the quench of $\rho_{gge1}$, we can define a
second GGE density matrix as
\begin{equation}\label{secondgge}
\rho_{gge2}= \frac{1}{Z_2}\prod_k \exp
\left(-\lambda_k^{(2)} \eta_k^\dagger(g_2) \eta_k(g_2)
\right),
\end{equation}
with the parameter $\lambda_k^{(2)} $ determined by the
condition $tr(\eta_k^\dagger(g_2)
    \eta_k(g_2) \rho_{gge2}) = tr(\eta_k^\dagger(g_2)
    \eta_k(g_2) \rho_{gge1})$, and $Z_2=\prod_k (1+ e^{-\lambda_k^{(2)}})$. The point is
that the basic correlator $G_{l,m}^{(2)}$ in (\ref{bmal3})
and the transverse magnetization in (\ref{transemag2}) can
be exactly reproduced by $\rho_{gge2}$. This is one more
support of the argument that $\rho_{gge1}$ can serve as a
pseudo-initial state to the second quench.

\textit{Expansion of hard-core bosons in a one dimensional
lattice}.---To make contact with previous works, the
scenario studied below is an extension of that in
Ref.~\cite{rigol07}. There are $N$ hard-core bosons and
there is a lattice of $M_{2}$ sites, which are numbered
from $1$ to $M_2$. Initially the $N$ bosons are confined to
the $M_0$ middle sites by hard-walls on the two sides and
the system is in the ground state, which is denoted as
$\psi_0$. At $t=0$, the hard-walls are suddenly moved
outward symmetrically so that now $M_{1}$ sites are
contained. The system then evolves and as found by Rigol
\textit{et al}. \cite{rigol07}, the GGE plays an important
role in the ensuing dynamics---the momentum distribution of
the bosons in its steady value is accurately captured by
the GGE density matrix $\Xi_{gge1}$ (see below). Our idea
is then at some time $t_1$, when the momentum distribution
has settled down to its steady value, to increase the
volume to $M_2$ sites and let the bosons expand once again.
The aim is to see whether the subsequent dynamics can be
accurately reproduced with the initial state (to the second
expansion) $\psi(t_1)$ replaced by $\Xi_{gge1}$. Note that
since the latter is time independent, this necessarily
requires that the subsequent dynamics be insensitive to the
specific value of $t_1$ as long as it is large enough to
belong to the steady regime.
\begin{figure}[tb]
\includegraphics[ width=0.35\textwidth, bb= 26   183   578   615]{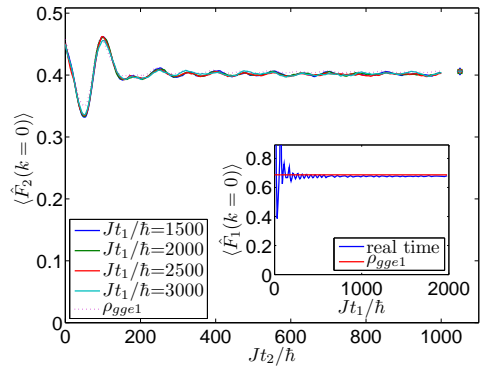}
\caption{(Color online) Evolution of the population on the
$k=0$ quasi-momentum state $\langle \hat{F}_2(k=0) \rangle$
after the second expansion. The parameters are
$(N,M_0,M_1,M_2)= (50,100,200,300)$. The dotted line
indicates the result with the ``initial'' state being the
(first) generalized Gibbs ensemble (GGE) density matrix
$\Xi_{gge1}$. Other lines correspond to results with the
``initial'' states being $\psi(t_1)$, with the value of
$t_1$ varied. The markers on the right ends of the lines
indicate the predicted values of the second GGEs. The
insert shows the time evolution of the population on the
$k=0$ quasi-momentum state $\langle \hat{F}_1(k=0) \rangle$
after the first expansion.\label{fig2}}
\end{figure}

In the intervals of $t\leq0$, $0< t < t_1$, and $t\geq
t_1$, the volume (number of sites) of the system is $M_0$,
$M_1$, and $M_2$, and thus the corresponding Hamiltonians
will be denoted as $H_0$, $H_1$, and $H_2$, respectively.
They are of the form $ H_i=-J\sum_{j=L_i}^{R_i-1}
(b_j^\dagger b_{j+1}+ b_{j+1}^\dagger b_j )$, $0\leq i \leq
2$. Here $J$ is the hopping strength, and $L_i=
(M_2-M_i)/2+1 $ and $R_i= (M_2+ M_i)/2$ denote the left-
and right-most sites accessible to the bosons,
respectively. The creation and annihilation operators
satisfy the usual bosonic commutation relations plus the
hard-core constraint $b_j^2 =b_j^{\dagger 2} =0$, so that
each site can be occupied by at most one boson. By using
the Jordan-Wigner transformation $(b_j^\dagger,b_j )=(
c_j^\dagger ,c_j ) \prod_{j'=1}^{j-1} e^{-i\pi
c_{j'}^\dagger c_{j'}}$, where $c_j$ ($c_j^\dagger$) is the
fermionic annihilation (creation) operator, $H_i$ is mapped
to a free fermion one, $H_i=-J\sum_{j=L_i}^{R_i-1}
(c_j^\dagger c_{j+1}+ c_{j+1}^\dagger c_j )$. This
Hamiltonian can be readily diagonalized as
$H_i=\sum_{q=1}^{M_i} \varepsilon_q^{(i)}
\eta^{(i)\dagger}_q \eta_q^{(i)}$, with
$\varepsilon_q^{(i)} = -2J \cos\left(\pi q/(M_i+1) \right)$
and $\eta_q^{(i)} =
\sqrt{\frac{2}{M_i+1}}\sum_{j=L_i}^{R_i} c_j \sin ( q\pi
(j-L_i+1)/(M_i+1) )$. The initial state is then simply a
Fermi-sea state $\psi_0=\prod_{q=1}^{N} \eta^{(0)\dagger}_q
|0\rangle$.

From $H_0$ to $H_1$, the wave function evolves as
$\psi(t_1)= e^{-i H_1 t_1/\hbar} \psi_0$, and the (first)
GGE density matrix is defined as $\Xi_{gge1} =
\frac{1}{\Theta_1} \exp [- \sum_{q=1}^{M_1}
\lambda_{q}^{(1)} c_{q}^{(1)\dagger} c_q^{(1)} ]$.
%\begin{equation}\label{gge1}
%    \Xi_{gge1} = \frac{1}{\Theta_1} \exp \left[- \sum_{q=1}^{M_1} \lambda_{q}^{(1)} c_{q}^{(1)\dagger} c_q^{(1)} \right].
%\end{equation}
Here the parameter $\lambda_q^{(1)}$ is determined by the
initial state, $tr( c_{q}^{(1)\dagger}
c_q^{(1)}\Xi_{gge1})= \langle \psi_0 | c_{q}^{(1)\dagger}
c_q^{(1)} | \psi_0 \rangle$,
%\begin{equation}\label{lambdaq}
%    \frac{1}{e^{\lambda_q^{(1)}} +1 }=tr( c_{q}^{(1)\dagger} c_q^{(1)}\Xi_{gge1})= \langle \psi_0 | c_{q}^{(1)\dagger}
%c_q^{(1)} | \psi_0 \rangle,
%\end{equation}
and $\Theta_1$ is a normalization factor (or partition
function). It is found in \cite{rigol07}, argued in
\cite{cazalilla11}, and verified in Fig.~\ref{fig2} below
that for $t_1$ large enough, the momentum distribution (or
populations on the quasi-momentum states, here
$k=-M_1/2,-M_1/2+1,\cdots,M_1/2-1$)
\begin{equation}\label{mom1}
    \hat{F}_{1} (k)  = \frac{1}{M_{1}}
    \sum_{j,j'=L_1}^{R_1} e^{-i 2\pi k (j-j')/M_{1}}
    b_{j'}^\dagger b_j
\end{equation}
with respect to $\psi(t_1)$ can be accurately reproduced by
using $\Xi_{gge1}$, i.e., $\langle \psi(t_1) | \hat{F}_1(k)
| \psi(t_1) \rangle\simeq tr(\hat{F}_1(k) \Xi_{gge1} )$.

Now from $H_1$ to $H_2$, the $H_2$-evolved wave function at
$t=t_1 + t_2$ is given by $\psi(t_1+t_2)=e^{-i H_2
t_2/\hbar} \psi(t_1)$. For our purpose, we replace the
``initial'' state $\psi(t_1)$ by $\Xi_{gge1}$ and define
the $H_2$-evolved GGE density matrix $\Xi_{gge1}(t_2)=
e^{-i H_2 t_2/\hbar} \Xi_{gge1} e^{i H_2 t_2/\hbar}$. We
then study the momentum distribution
($k=-M_2/2,-M_2/2+1,\cdots,M_2/2-1$)
\begin{equation}\label{mom2}
    \hat{F}_{2} (k)  = \frac{1}{M_{2}}
    \sum_{j,j'=L_2}^{R_2} e^{-i 2\pi k (j-j')/M_{2}}
    b_{j'}^\dagger b_j
\end{equation}
with respect to $\psi(t_1 + t_2)$ and $\Xi_{gge1}(t_2)$.
The results are shown in Fig.~\ref{fig2}.

In the insert of Fig.~\ref{fig2}, we see that after the
first expansion, the population on the $k=0$ quasi-momentum
state $\langle \psi(t_1) |\hat{F}_{1} (k=0) | \psi(t_1)
\rangle $ relaxes to the steady value predicted by the GGE
density matrix $ \Xi_{gge1}$ eventually. This proves the
predictive power of the GGE after the first expansion. What
Fig.~\ref{fig2} highlights is that, if the time of the
second expansion $t_1$ is chosen to belong to the steady
regime, the later evolution of the population on the $k=0$
quasi-momentum state $\langle \psi(t_1+t_2) |\hat{F}_{2}
(k=0) | \psi(t_1+t_2) \rangle $ can be accurately
reproduced by $tr(\hat{F}_{2} (k=0)  \Xi_{gge1}(t_2))$.
Their lines coincide with each other not only in the
asymptotic limit but even on details during the transitory
period. Note that since the latter is independent of $t_1$,
this necessarily implies that the former is insensitive to
the value of $t_1$, as is indeed the case. Overall,
Fig.~\ref{fig2} is a remarkable demonstration of the fact
that the GGE density matrix $\Xi_{gge1}$ shares with the
relaxed state $\psi(t_1)$ not only the value of the
momentum distribution, but also the response to a second
quench. Or in the perspective of the state $\psi(t_1)$, it
has relaxed to be virtually indistinguishable from the GGE
state $\Xi_{gge1}$, neither by static nor dynamical
criterions.

In Fig.~\ref{fig2}, we have also studied whether the steady
value of $\hat{F}_{2} (k=0) $ after the second quench can
be described by a second GGE density matrix $\Xi_{gge2}$,
which is defined as $\Xi_{gge2} = \frac{1}{\Theta_2} \exp
[- \sum_{q=1}^{M_2} \lambda_{q}^{(2)} c_{q}^{(2)\dagger}
c_q^{(2)} ]$,
%\begin{equation}\label{rhogge2}
%   \Xi_{gge2} = \frac{1}{\Theta_2} \exp \left[-
%\sum_{q=1}^{M_2} \lambda_{q}^{(2)} c_{q}^{(2)\dagger}
%c_q^{(2)} \right],
%\end{equation}
with the parameter $\lambda_{q}^{(2)} $ determined by the
condition $tr(c_{q}^{(2)\dagger} c_q^{(2)} \Xi_{gge2}
)=tr(c_{q}^{(2)\dagger} c_q^{(2)} \Xi_{gge1} )$ or $
tr(c_{q}^{(2)\dagger} c_q^{(2)} \Xi_{gge2} )= \langle
\psi(t_1)| c_{q}^{(2)\dagger} c_q^{(2)} | \psi(t_1)
\rangle$ depending on whether the ``initial'' state is
$\Xi_{gge1}$ or $\psi(t_1)$. The result is that the second
GGEs do predict the steady values correctly; moreover, they
agree with each other very well. This is one more evidence
that the relaxed wave function $\psi(t_1)$ is virtually
indistinguishable from the GGE $\Xi_{gge1}$.

In summary, we have investigated and verified the relevance
of the GGEs in the dynamical response of the two integrable
models of transverse Ising model and one-dimensional
hard-core bosons. Once having relaxed to have its
properties correctly predicted by the GGE, the system
behaves as if it were indeed in the GGE state---its
response to the second quench can be accurately reproduced
by the GGE even to details. On one hand, this result is a
welcome complement to previously established result that
the GGEs are relevant in predicting the static properties
of the systems after the first quench. The two now combine
to present a more complete story of the GGE and beckon more
confidence on it. On the other hand, this result also gives
us a sense of ``dynamical typicality'' \cite{gemmer}, which
is also observed in the (non-integrable) Bose-Hubbard model
previously \cite{zjm}. Finally, though here we have been
dealing with integrable systems only, a lesson may also be
drawn for non-integrable systems. A closed non-integrable
system might well be a pure state yet virtually
indistinguishable neither by static nor by dynamic
criterions from a canonical ensemble.

We acknowledge Institute of Physics, CAS for funding.

\end{document}